\documentclass[aps, prd, 10pt, twocolumn, tightenlines, notitlepage, superscriptaddress, nofootinbib, preprintnumbers, floatfix, showkeys]{revtex4-2}
\usepackage{soul}
\usepackage{upgreek}
\usepackage[normalem]{ulem}
\usepackage{amstext}
\usepackage{amssymb}
\usepackage{amsmath}
\usepackage{graphicx}
\usepackage{url}
\usepackage{color}
\usepackage{ulem}
\usepackage[utf8]{inputenc}
\pdfoutput=1
\usepackage{textcomp}
\usepackage{booktabs}
\usepackage{booktabs,siunitx,array,threeparttable}
\usepackage[T1]{fontenc}
\usepackage{appendix}

\usepackage{epsfig,amsfonts,mathrsfs,amsmath,amssymb,graphicx,color,slashed,multirow}
\usepackage{amsmath,latexsym,amssymb,graphicx,slashed,color,enumerate,url,cancel,gensymb}
\usepackage{textcomp}

\usepackage[x11names]{xcolor}
\usepackage[colorlinks]{hyperref}
\usepackage{orcidlink}

\usepackage[paperwidth=210mm,paperheight=297mm,centering,hmargin=2.5cm,vmargin=2.5cm]{geometry}
\usepackage{tikz-feynman}
\tikzfeynmanset{compat=1.1.0}
\tikzfeynmanset{warn luatex=false}
\usetikzlibrary{patterns,decorations.pathreplacing}
\usetikzlibrary{decorations.pathmorphing}
\usetikzlibrary{decorations.markings}
\usetikzlibrary{arrows,shapes}
\usetikzlibrary{shapes.misc}
\tikzset{
  gluon/.style={decorate, draw=black,
    decoration={coil,amplitude=4pt, segment length=4pt,aspect=0.7}} 
}
\tikzset{
  photon/.style={decorate, decoration={snake}},
}

\usepackage{textgreek} 
\usepackage{booktabs} 

\hypersetup{colorlinks,citecolor= nicered,linkcolor= blue}
\definecolor{nicered}{rgb}{0.7,0.1,0.1}
\definecolor{nicegreen}{rgb}{0.1,0.5,0.1}

\def\cevns{CE$\upnu$NS}

\newcommand{\ETmiss}{{E_{\rm{T}}^{\rm{miss}}}}

\definecolor{myblue}{cmyk}{0.65, 0.37, 0.0, 0.19}

\definecolor{blue(ncs)}{rgb}{0.0, 0.53, 0.74}

\definecolor{mypurple}{cmyk}{.44, 1, 0, 0}

\AtBeginDocument{\hypersetup{citecolor=nicered,linkcolor=nicered,urlcolor=nicered}}

\begin{document}

\title{\texorpdfstring{\Large
Neutrino nonstandard interactions:\\
Confronting COHERENT and LHC data}{Neutrino nonstandard interactions: Confronting COHERENT and LHC data}}

\author{Víctor Martín Lozano~\orcidlink{0000-0002-9601-0347} }\email{victor.lozano@ific.uv.es}
\affiliation{Instituto de F\'isica Corpuscular (IFIC), CSIC‐Universitat de Val\`encia, Parc Cient\'ific UV C/ Catedr\'atico Jos\'e Beltr\'an, 2 E-46980 Paterna (Valencia), Spain}
\affiliation{Departament de F\'isica Teòrica, Universitat de Val\`{e}ncia, 46100 Burjassot, Spain}
\author{G. Sanchez Garcia~\orcidlink{0000-0003-1830-2325}}\email{gsanchez@ific.uv.es}
\affiliation{Instituto de F\'isica Corpuscular (IFIC), CSIC‐Universitat de Val\`encia, Parc Cient\'ific UV C/ Catedr\'atico Jos\'e Beltr\'an, 2 E-46980 Paterna (Valencia), Spain}
\affiliation{Departament de F\'isica Teòrica, Universitat de Val\`{e}ncia, 46100 Burjassot, Spain}
\author{Adri\'an Terrones~\orcidlink{0000-0003-1636-3335}}\email{adrian.terrones@ific.uv.es}
\affiliation{Instituto de F\'isica Corpuscular (IFIC), CSIC‐Universitat de Val\`encia, Parc Cient\'ific UV C/ Catedr\'atico Jos\'e Beltr\'an, 2 E-46980 Paterna (Valencia), Spain}


\begin{abstract}
We study the complementarity between COHERENT and LHC searches in testing neutrino nonstandard interactions (NSIs) through the completion of the effective field theory approach within a $Z'$ simplified model. Our results show that LHC bounds are strongly dependent on the $Z'$ mass, with relatively large masses excluding regions in the parameter space that are allowed by COHERENT data and its future expectations. We demonstrate that the combination of low- and high-energy experiments results in a viable approach to break NSI degeneracies within the context of simplified models.
\end{abstract}

\maketitle

\section{Introduction}

The neutrino is one of the most enigmatic particles in the Standard Model (SM), a theory in which they are considered to be massless. However, the SM is not complete, and the discovery of neutrino oscillations \cite{Kajita:2016cak, RevModPhys.88.030502} represented an unambiguous proof that, indeed, they have a small mass.
The mechanism responsible for the origin of neutrino masses is still unknown and motivates the study of new interactions that lie beyond the SM.
A plethora of models have been proposed to explain the smallness of neutrino masses by different mechanisms, either by considering extra neutrinos~\cite{Minkowski:1977sc, Mohapatra:1979ia}, expanding the scalar sector~\cite{Schechter:1981cv, Schechter:1980gr, Zee:1980ai, Zee:1985id,Tao:1996vb, Ma:2006km}, introducing new fermions~\cite{Foot:1988aq}, or resorting to new symmetries~\cite{deAnda:2018ecu,Nomura:2019yft,Ding:2019zxk,Okada:2019mjf,ChuliaCentelles:2022ogm,CentellesChulia:2023osj}. Regardless of its origin, the underlying new physics is expected to manifest itself as deviations from the SM predictions at neutrino and accelerator experiments. Interestingly, the energy range at which new physics effects can be present covers from a few MeV, passing through the TeV scale, up to energies at the Planck scale. Experimentally, the MeV range can be widely tested in direct detection neutrino experiments, with sources including reactors~\cite{Leitner:2011aa, Ohlsson:2008gx, Khan:2013hva}, pion decay-at-rest sources~\cite{Giunti:2019xpr, DeRomeri:2022twg}, and solar neutrinos~\cite{Miranda:2004nb, Miranda:2015dra, Gonzalez-Garcia:2013usa}. At this energy level, where the momentum transfer is negligible, effective theories can be used to test new physics scenarios in a general way, with nonstandard interactions (NSIs) as the most widely used formalism~\cite{Wolfenstein:1977ue}. On the other hand, in the TeV range the momentum transfer is non-negligible, and effective theories may lose their validity. In this case, the degrees of freedom from the underlying theory must be taken into account and will manifest as decay products of new mediators of the theory. Fortunately, one can make use of simplified models that reproduce NSI-type interactions like the ones from low-energy ranges. In this way, experiments at MeV and TeV scales can be used to test NSIs across different regimes~\cite{Barger:1991ae, Grossman:1995wx}.

The observation of coherent elastic neutrino-nucleus scattering (\cevns) by the COHERENT Collaboration~\cite{COHERENT:2017ipa} has opened the gate to explore the implications of new physics in the neutrino sector at low energies. In particular, data from this process have been used to constrain NSIs, finding degeneracies in the parameter space that can be slightly broken when combining information from the same process but using different scattering target materials.
On the other hand, colliders can be used to understand the nature of NSIs in high-energy regimes~\cite{Berezhiani:2001rs,Barranco:2007ej,Davidson:2011kr,Friedland:2011za,BuarqueFranzosi:2015qil,Choudhury:2018xsm,Han:2019zkz,Babu:2020nna,Liu:2020emq,Cheung:2021tmx,Yue:2022eac,Jana:2023ogd,Escrihuela:2023sfb}. 
However, in this case, the effective field approach is no longer valid, and one should rely on an underlying theory. One of the simplest models to address these types of interactions is the use of simplified $Z'$ models that can come from UV completions~\cite{Fox:2011pm,Busoni:2013lha,Buchmueller:2013dya,Goncalves:2016iyg}.

In this work, we will be interested in confronting neutrino NSIs from both the low- and high-energy regimes using the available experimental data from COHERENT and LHC. For this purpose, we will make use of an effective field theory (EFT) description of NSIs that matches with a simplified model containing a new vector mediator $Z'$ as its high-energy completion. As we will see, the study of NSIs at the LHC is sensitive to both vector and axial currents, with the disadvantage that neutrino flavors in this case are indistinguishable. In that sense, LHC and \cevns~ searches can be complementary, leading to a better understanding of NSIs at high energies and breaking degeneracies appearing in neutrino scattering results.

This work is organized as follows. In Sec. \ref{sec:formalism} we build up the formalism of NSIs and its correspondence to the simplified $Z'$ models. In Sec.~\ref{sec:COHLHC} we describe the analysis performed for COHERENT and LHC data to study their sensitivity to NSIs. In Sec.~\ref{sec:results} we present our results over different scenarios and, finally, we give our conclusions in Sec.~\ref{sec:conclusions}.

\section{Formalism}
\label{sec:formalism}

The concept of nonstandard interactions is generally used to describe a low-energy parametrization that accounts for charged and neutral currents coming from physics beyond the SM in the leptonic sector. Under this formalism, new physics effects are encoded in terms of six-dimensional operators that are seen as an effective field theory of the SM below the electroweak scale. In the following, we will be only interested in the study of neutral current NSIs (NC-NSIs), for which the corresponding operator has the form
\begin{equation}
    \mathcal{L}_{{\rm NSI}}^{\rm NC}= -2 \sqrt{2}G_F \varepsilon_{\alpha\beta}^{fC} (\bar{\nu}_\alpha \gamma^\mu P_L \nu_\beta)(\bar{f} \gamma_\mu P_C f) \,,
    \label{eq:NSI}
\end{equation}
where $\alpha,\beta= e, \mu, \tau$ denote the different neutrino flavors, $f=u,d,e$ refers to matter fermion fields, and $C=L,R$ stands for the left- and right-handed chiralities. The operators in Eq. \eqref{eq:NSI} are scaled by the Fermi constant, $G_F$, and the strength of the interactions is given by the parameters $\varepsilon_{\alpha\beta}^{fC}$, which are widely known as NSI parameters in the literature. Since we want NSI effects to be subdominant with respect to weak interactions, NSI parameters are expected to be of order $\leq \mathcal{O}(1)$. For the sake of simplicity, throughout this work we will only focus on the presence of vector interactions for the NC-NSIs, which are defined as $\varepsilon^{qV}_{\alpha\beta}\equiv \varepsilon_{\alpha\beta}^{qL} + \varepsilon_{\alpha\beta}^{qR}$.

Theoretically, there are different ways to generate the operators defined in Eq. \eqref{eq:NSI}. For instance, they may result from a gauge-invariant six-dimensional operator before electroweak symmetry breaking,
\begin{equation}
\frac{c^{QQ}_{LL}}{\Lambda^2}(\bar{L}_\alpha \gamma_\mu L_\beta)(\bar{Q}\gamma^\mu Q)\,,
\label{eq:6dim}
\end{equation}
where $L$ denotes the SM lepton doublets, and $Q=Q_L, q_R$ are the SM quark doublets and singlets, respectively, $c^{QQ}_{LL}$ is the operator's Wilson coefficient, and $\Lambda$ accounts for the new physics scale. The singlet structure gives $\bar{L}_\alpha \gamma_\mu L_\beta=\bar{\nu}_\alpha \gamma_\mu \nu_\beta + \bar{\ell}_\alpha \gamma_\mu \ell_\beta$ and it is related to the NSI structure as
\begin{equation}
    \varepsilon_{\alpha\beta}=\frac{c_{LL}^{QQ}}{2\sqrt{2}G_F\Lambda^2}\,.
\end{equation}

Another possible origin for the Lagrangian in Eq.~\eqref{eq:NSI} comes from higher-dimensional operators that involve the coupling of the scalar sector with the lepton fields. For instance, the eight-dimensional operator
\begin{equation}
    \frac{c_8}{\Lambda_8^4}(\bar{H}\bar{L}_\alpha\gamma_\mu H L_\beta)(\bar{Q}\gamma^\mu Q)\,,
\end{equation}
 where $H$ is the Higgs doublet, can give rise to NSI couplings as studied in Ref.~\cite{Davidson:2011kr}, where the NSI coefficients would be given by $\varepsilon_{\alpha\beta} = c_8v^2/(2\sqrt{2}G_F\Lambda_8^4)$.

The previous description in terms of an effective field theory is convenient in different low-energy experiments where the momentum transfer is of the order of a few MeV~\cite{Breso-Pla:2023tnz}. However, at collider experiments like the LHC, the center-of-mass energy $\sqrt{s}$, of the different processes, can reach values of the order of the TeV scale. In this regime, the validity of the EFT is not guaranteed, and one should make use of an alternative formalism to generate the necessary operators to get NSIs. There exist several complete models that can accommodate these kinds of operators at low energy, such as leptoquarks or R-parity violating supersymmetry~\cite{Barranco:2007tz,Billard:2018jnl,Crivellin:2021bkd,Calabrese:2022mnp,DeRomeri:2023cjt}. Another option is to assume a simplified model with a $Z'$ mediator having an arbitrary mass $m_{Z'}$~\cite{Fox:2011pm,Busoni:2013lha,Buchmueller:2013dya,Goncalves:2016iyg}. Then, NSIs are originated through the Lagrangian term\footnote{Note that NSIs can also be generated through, for instance, scalar mediators. However, in that case, operators would either depend on field  derivatives or the analysis would require searches that are beyond the scope of this work.}
\begin{equation}
    \mathcal{L}_{\rm NSI}^{\rm simp}=(g_\nu^{\alpha\beta}\bar{\nu}_\alpha \gamma^\mu P_L\nu_\beta + g_{q_i}^C \bar{q}_i \gamma^\mu P_C q_i)Z^\prime_{\mu}\,,
    \label{eq:simnsineu}
\end{equation}
where $\alpha, \beta = e, \mu, \tau$ are flavor indices, $i=u,d$, and $P_C$ stands for the left and right projectors ($C=L,R$), while the constants $g_\nu^{\alpha\beta}$ and $g_{q_i}^C$ are the couplings of the $Z'$ to neutrinos and quarks, respectively. The Lagrangian in Eq.~\eqref{eq:simnsineu} gives rise to neutral current interactions where we can identify NSI couplings as
\begin{equation}
    \varepsilon_{\alpha \beta}^{V}=\frac{g_{\nu}^{\alpha, \beta}g_{u,d}^V}{2\sqrt{2}G_F m_{Z^\prime}^2}\,,
\end{equation}
where we have used $g_{u,d}^V=g_{u,d}^L + g_{u,d}^R$ and, for the sake of simplicity, we assumed $g_{u,d}^A = 0$.\footnote{Although LHC is sensitive to axial-vector couplings, experiments involving \cevns\, such as COHERENT are rather insensitive to them, so in order to simply compare both results we keep only vector interactions. However, other scattering experiments can be sensitive to axial NSIs~\cite{Escrihuela:2011cf,Abbaslu:2023vqk}.} If, on the other hand, new physics comes from a six-dimensional operator as in Eq.~\eqref{eq:6dim}, then NSIs can be generated by the simplified Lagrangian,
\begin{multline}
    \mathcal{L}_{\rm NSI}^{\rm simp}=[g_\ell^{\alpha\beta}(\bar{\nu}_\alpha \gamma^\mu P_L\nu_\mu + \bar{\ell}_\alpha \gamma^\mu P_L\ell_\beta) \\ + g_{q_i}^C \bar{q}_i \gamma^\mu P_C q_i]Z^\prime_{\mu}\,,
    \label{eq:simnsileptons}
\end{multline}
where $\ell=e,\mu,\tau$ and we have assumed that neutral and charged leptons have the same coupling $g_\ell^{\alpha\beta}$ to the $Z'$ mediator. Note that, in this approach, those terms with $\alpha \neq \beta$ induce lepton flavor changing currents, and therefore are subject to severe constraints~\cite{Bergmann:1998ft}. Then, in the following, we will assume only NSIs with $\alpha=\beta$ to be nonzero. Under this scenario, the NSI couplings arising from the simplified model are
\begin{equation}
   \varepsilon_{\alpha \beta}^{V}=\frac{g_{\ell}^{\alpha, \beta}g_{u,d}^V}{2\sqrt{2}G_F m_{Z^\prime}^2}\,.
\end{equation}

As a last remark, if $\varepsilon_{\alpha \beta}^{u,d}$ is arbitrarily large for a given $m_{Z'}$ the EFT approach is no longer valid at the LHC. Following Ref.~\cite{Babu:2020nna}, this limit can be obtained when the total width of the $Z'$ is larger than the partial widths into leptons and quarks. Under this condition, one obtains
\begin{equation}
    |\tilde{\varepsilon}|\leq \frac{\sqrt{3}\pi}{\sqrt{n}\,G_Fm_{Z^\prime}^2}\frac{\Gamma_{Z^\prime}}{m_{Z^\prime}}\,,
    \label{eq:pert}
\end{equation}
where $\Gamma_{Z^\prime}$ is the decay width of the $Z^\prime$, $n$ is the number of quark flavors below the threshold for the $Z^\prime$ to decay, and $|\tilde{\varepsilon}|^2=\sum_{\alpha\beta}|\varepsilon_{\alpha\beta}|^2$. Assuming a greater value of $\varepsilon$ would imply a larger value of the width $\Gamma_{Z'}$, leading to nonperturbativity and the equivalence into an effective field theory not being valid.

\section{Low- and High-Energy Constraints}
\label{sec:COHLHC}

\subsection{COHERENT}

In the low-energy regime, NSIs can be probed through the interaction of neutrinos with nuclei. That is the case of \cevns, a neutral current process predicted by the SM~\cite{PhysRevD.9.1389}, in which a neutrino interacts with an entire nucleus. The coherence condition for this process requires the incoming neutrino energy to be at most a few tens of MeV, making~\cevns\ a suitable tool for testing NSIs at low energies. Within the SM, the~\cevns\ differential cross section is given by
\begin{equation}
\label{eq:cross:SM}
\left (  \frac{\mathrm{d}\sigma}{\mathrm{d}T} \right )_{\textrm{SM}} = \frac{G_F^2M}{\pi}\left(1-\frac{MT}{2E_{\nu}^2}\right)\,\left(Q_{W}^V\right)^2\,,
\end{equation}
where $M$ is the mass of the nucleus, $E_\nu$ is the energy carried by the neutrino, $T$ is the nuclear recoil energy acquired by the nucleus as a result of the interaction, and $Q_W^V$ is the SM weak charge, given by
\begin{equation}\label{eq:qweak}
    \left(Q_{W}^V\right)^2  = \left[ZF_Z(q^2)\,g_V^p + NF_N(q^2)\,g_V^n \right]^2\,,
\end{equation}
with $Z$ and $N$ the number of protons and neutrons of the target materials, respectively, and $F_{Z,N}$ the corresponding nuclear form factor~\cite{Lewin:1995rx, Klein:1999qj}. After introducing diagonal NSIs, the SM cross section is modified by a redefinition of the weak charge, which now reads~\cite{Barranco:2005yy}
\begin{multline}
\left(Q_{W,\alpha}^{V,\textrm{NSI}}\right)^2  = \big[ZF_Z(q^2)\left(g_V^p + 2\varepsilon_{\alpha\alpha}^{uV}+\varepsilon_{\alpha\alpha}^{dV}\right) \\
+ NF_N(q^2)\left(g_V^n + \varepsilon_{\alpha\alpha}^{uV}+2\varepsilon_{\alpha\alpha}^{dV}\right)\big]^2\,,
 \label{eq:weak:charge:NSI}
\end{multline}
allowing for~\cevns\ processes to test different NSIs. On the experimental side, this process has been detected by the COHERENT Collaboration using neutrinos from a pion decay-at-rest source, and with three different detection technologies based on CsI~\cite{COHERENT:2017ipa, COHERENT:2021xmm}, LAr~\cite{COHERENT:2020iec} and Ge~\cite{COHERENT:2024axu}. Additionally, there is strong evidence of its observation from $^{8}$B solar neutrinos in Xe-based detectors by the PandaX-4T~\cite{PandaX:2024muv} and XenonNT~\cite{XENON:2024ijk} direct detection experiments. Moreover, the CONUS+ Collaboration has recently reported an observation using reactor antineutrino sources~\cite{Ackermann:2025obx}. The impact of these measurements on diagonal NSI bounds has been studied in a variety of references, being spallation sources sensitive to both $\varepsilon_{ee}^{qV}$ and $\varepsilon_{\mu\mu}^{qV}$~\cite{DeRomeri:2022twg, Coloma:2017ncl}, reactor sources only to $\varepsilon_{ee}^{qV}$~\cite{DeRomeri:2025csu, Alpizar-Venegas:2025wor}, and solar neutrinos the only source to probe $\varepsilon_{\tau\tau}^{qV}$~\cite{ AristizabalSierra:2024nwf}. Among all \cevns\ analysis in the literature, the most stringent NSI bounds come from COHERENT data. Hence, we focus on this experiment to illustrate the complementarity between low- and high-energy ranges when constraining diagonal NSIs. Here we present the bounds that result from a combined analysis of the CsI and LAr detectors,\footnote{Given the low statistics of the Ge COHERENT detector, we do not expect adding these data to the combined analysis will result in a significant change.} taking the constraints obtained in Ref.~\cite{DeRomeri:2022twg} when possible and following an identical $\chi^2$ procedure used in the same reference otherwise. 

\subsection{LHC}
NSIs that result from simplified models can be tested at hadron colliders like the LHC, where the $Z^\prime$ can be produced through quark annihilation. Here we consider two different scenarios, depending on whether, in the lepton sector, the $Z'$ couples only to neutrinos or to both neutrinos and charged leptons. The case in which the $Z'$ only couples to neutrinos is given by the Lagrangian in Eq.~\eqref{eq:simnsineu}. After being produced, in this case, the $Z'$ can decay into a pair of neutrinos, which would not be directly detectable since neutrinos are invisible to the LHC. However, we can rely on the production of a jet as initial state radiation to tag the transverse missing energy and probe the new physics contributions through the process $pp\to j+ \ETmiss$, where $j$ stands for the initial state radiation jet and $\ETmiss$ for the missing transverse energy. This approach has been shown to be relevant to set constraints on NSI couplings~\cite{Friedland:2011za,BuarqueFranzosi:2015qil,Choudhury:2018xsm,Babu:2020nna,Liu:2020emq,Cheung:2021tmx}. As the $Z'$ also couples to quarks, in this case, it can also decay into a pair of jets after being produced, generating a dijet signal in addition to the monojet one. If, on the other hand, we adopt the scheme of the simplified model of Eq.~\eqref{eq:simnsileptons}, then the $Z'$ also couples to charged leptons and in this case dilepton signals can be very powerful to constrain NSIs~\cite{Choudhury:2018xsm,Han:2019zkz}. In either case, the probability of the $Z'$ to decay to the different fermions through vector currents is given by its decay width,
\begin{equation}
    \Gamma_{Z'}^{i}=\frac{N_C^i (g_i^V)^2m_{Z'}}{12\pi}\sqrt{1-\frac{4m_i^2}{m_{Z'}^2}}\left(1+\frac{2m_i^2}{m_{Z'}^2}\right)\,,
\end{equation}
where $N_C^i=3$ for quarks and $N_C^i=3$ for leptons, and $i$ denotes all fermions appearing in Eq.~\eqref{eq:simnsileptons}.

To test the sensitivity of the LHC to NSI parameters we use {\tt $Z^\prime$-explorer}~\cite{Alvarez:2020yim,Lozano:2021zbu}, a software tool that tests $Z^\prime$ models against LHC results. In particular, {\tt $Z^\prime$-explorer} contains the databases of the ATLAS and CMS dijet~\cite{ATLAS:2019fgd}, $e^+e^-$~\cite{CMS:2021ctt}, $\mu^+\mu^-$~\cite{CMS:2021ctt}, $\tau^+\tau^-$~\cite{ATLAS:2017eiz}, and monojet~\cite{ATLAS:2021kxv} channels. To test a specific model, {\tt $Z^\prime$-explorer} takes as an input the $Z'$ mass $m_{Z'}$ and its couplings to the rest of the particles and computes the predicted cross section for all channels. Then, for each point, the signal strength of every decay channel is determined as
\begin{equation}
    \mathcal{S}=\frac{\sigma_{\rm pred}}{\sigma_{\rm lim}^{95}}\,,
\end{equation}
where $\sigma_{\rm pred}$ denotes the predicted cross section $\times$ branching ratio $\times$ acceptance, and $\sigma_{\rm lim}^{95}$ is the corresponding predicted experimental upper limit at the 95\% confidence level (CL). If the result of the signal strength for a given point is $\mathcal{S}>1$ for at least one channel, then it is considered as excluded. However, if $\mathcal{S} < 1$ for a given point in all channels, then it is not currently excluded, and the channel with the largest strength is considered the most sensitive to exclude it in the future.

\section{NSI bounds from COHERENT and LHC}
\label{sec:results}

In this section, we present the different bounds obtained from low- and high-energy experiments. We discuss two different scenarios: one assuming NSIs to only be present in the neutrino sector, and another one by also including NSIs in the charged lepton sector.

\subsection{NSIs in the neutrino sector}
We begin by assuming that the only leptons participating in new interactions are neutrinos, which corresponds to the scenario described by the Lagrangian in Eq.~\eqref{eq:simnsineu}. Under this assumption, the only testable signals at the LHC would be mono- and dijet signatures.

\begin{figure}[t]
	\centering
        \includegraphics[width=.49\textwidth]{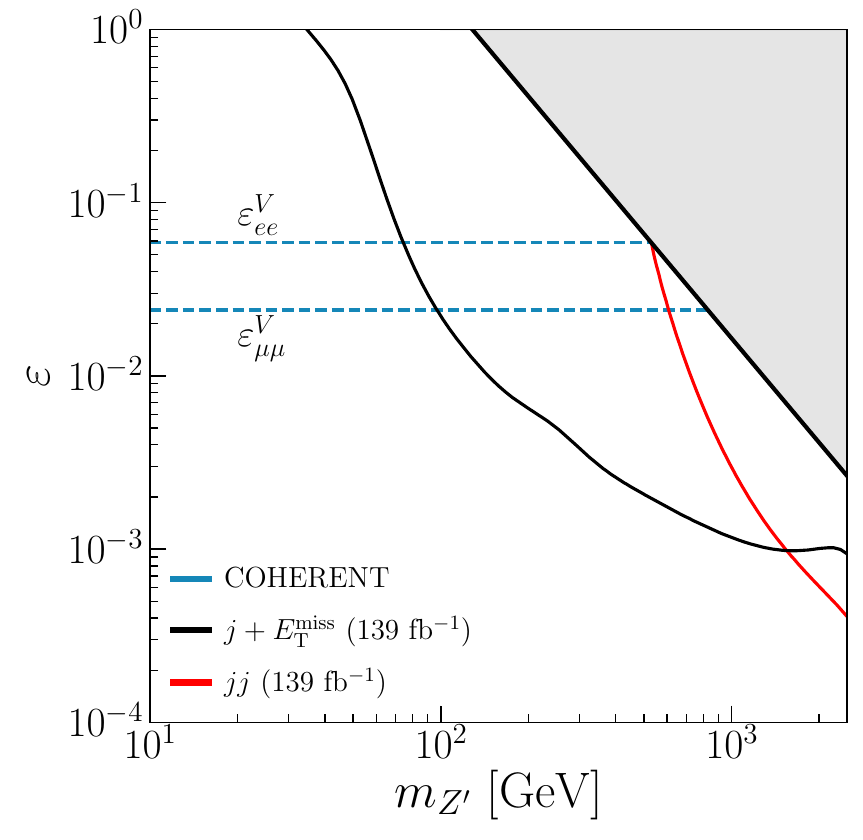}\hspace{1cm}
        \includegraphics[width=.49\textwidth]{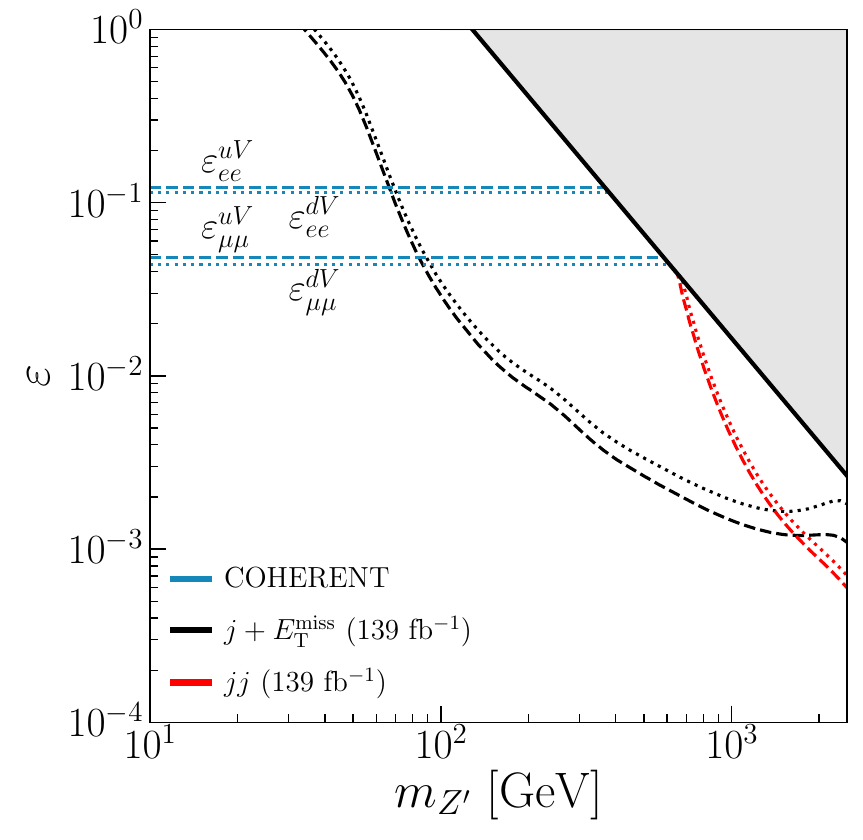}
	\caption{Constraints on NSI coupling, $\varepsilon$, as a function of $m_{Z'}$ from LHC and COHERENT data, for the cases where the $Z'$ couples equally to up and down quarks (top) and where it couples exclusively to either up or down quarks (bottom). In both scenarios, the $Z'$ interacts only with neutrino leptons . Mono- and dijet results are shown as black and red areas, respectively, while bounds from COHERENT data are depicted as horizontal lines. The gray region in the top-right corner corresponds to the parameter space where the validity of the EFT approach, given by Eq.~\eqref{eq:pert}, is not fulfilled.} 
	\label{fig:epsilonmz}
\end{figure}
 We show in Fig.~\ref{fig:epsilonmz} the LHC sensitivity to neutrino nonstandard interaction parameter $\varepsilon$ (see below), as a function of $m_{Z^\prime}$ for a fixed value of $g_\nu$, depicting the exclusion limits for mono- and dijet signals as the black and red regions, respectively. For reference, we also indicate as a light gray area the region where the effective field theory is inconsistent according to the criterion in Eq.~\eqref{eq:pert}. In the top panel of the figure, we have assumed that the $Z'$ couples to only one neutrino flavor and with the same strength to up and down quarks. Under these assumptions, the NSI bounds will be the same regardless of the neutrino flavor and quark type, so we can define
\begin{equation}
    \varepsilon=\varepsilon^{V}_{\alpha \alpha}=\frac{g_{\nu}^{\alpha, \alpha}g_{q}^V}{2\sqrt{2}G_F m_{Z^\prime}^2},
\end{equation}
for $\alpha=e,\mu$. As we can see from the figure, the LHC searches are sensitive to values of $\varepsilon$ smaller than 1 when $m_{Z^\prime}>~40$~GeV, becoming more sensitive for masses around 100 GeV, where LHC is able to constrain NSIs below $\varepsilon\simeq 10^{-2}$. Once the mass of the $Z'$ reaches $m_{Z'}\simeq$ 1--2 TeV the sensitivity of the search stops at $\varepsilon \simeq 10^{-4}$. Notice that the dijet search is only sensitive from masses starting at $m_{Z'}\simeq 500$ GeV, reaching values of $\varepsilon < 3\times10^{-4}$ for masses greater than $m_{Z'} >1$ TeV. One of the main reasons LHC searches are not sensitive to low values of $\varepsilon$ when $m_{Z'}$ has lower values is that, even if these analyses can constrain the couplings appearing in Eq.~\eqref{eq:simnsineu} for low masses, the mass in the denominator of Eq.~\eqref{eq:NSI} tends to increase the value of $\varepsilon$ even for low values of $g_{\nu}$. Regarding the low-energy regime, we provide as dashed lines on the same figure the 95\% C.L. on NSIs obtained through the analysis of COHERENT CsI+LAr data, where we denote $\varepsilon_{\alpha\alpha}^V\equiv \varepsilon_{\alpha\alpha}^{uV}=\varepsilon_{\alpha\alpha}^{dV}$, since we are assuming that neutrinos couple to both quark types with the same strength. Notice that, in contrast to LHC, the NSI constraints from COHERENT are sensitive to the neutrino flavor. Interestingly, the constraints for electron NSIs, $\varepsilon_{ee}^V$, are only competitive with LHC for $Z'$ masses below 70 GeV. On the other hand, the sensitivity of COHERENT to the muon NSIs, $\varepsilon_{\mu\mu}^V$, is stronger than the one to electrons. This allows one to impose bounds of $\varepsilon\simeq 2.5\times 10^{-2}$ up to masses $m_{Z'}\lesssim 90$ GeV. After this mass value, monojet searches are more robust and can constrain NSIs up to $10^{-3}$. However, dijet searches are more sensitive for $Z'$ masses satisfying $m_{Z^\prime}>1500$ GeV, showing the complementarity between the low- and high-energy processes.
 
In our previous analysis, it has been assumed that up and down quarks couple with the same strength to neutrinos. However, it is interesting to explore scenarios where neutrinos only couple to one quark flavor. We study this scenario in the bottom panel of Fig.~\ref{fig:epsilonmz}, where upper limits from mono- and dijet searches are shown as black and red lines, respectively. The up quark scenario is depicted as dashed lines, while the down quark one is shown as dotted lines. In general, we see that the limits for the case where neutrinos only couple to up quarks are stronger than those where they couple to down quarks. This is mainly because, as a proton collider, the probability of an up-type quark collision at the LHC is larger than one of a down-type. To see the complementarity with \cevns, we show again as horizontal lines the limits obtained from COHERENT data at 95\% C.L., which are consistent with the bounds computed in Ref. \cite{DeRomeri:2022twg}. The limits coming from up quarks are shown as dashed lines in the figure, while the dotted lines correspond to down quarks. We can notice that the sensitivity of the COHERENT limits depends strongly on the neutrino flavor rather than on the quark type, with the difference between the up and down bounds explained by the larger presence of neutrons than protons in CsI and Ar nuclei. As in the previous case, limits from electron neutrino NSIs are weaker than those coming from muons, the latter being more robust than LHC for masses $m_{Z'}\lesssim 90$ GeV. For masses greater than this value, the monojet starts to be the most sensitive observable that constrains NSIs and up to $m_{Z'}\simeq$1.5 TeV, where the dijet search becomes dominant.

\begin{figure*}[t]
\begin{minipage}[t]{0.90\textwidth}
    \includegraphics[width=\linewidth]{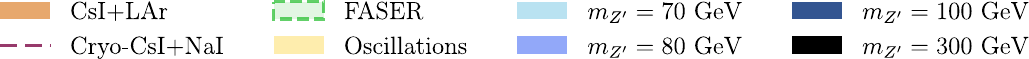}
\end{minipage}

\begin{minipage}[t]{0.48\textwidth}
  \includegraphics[width=\linewidth]{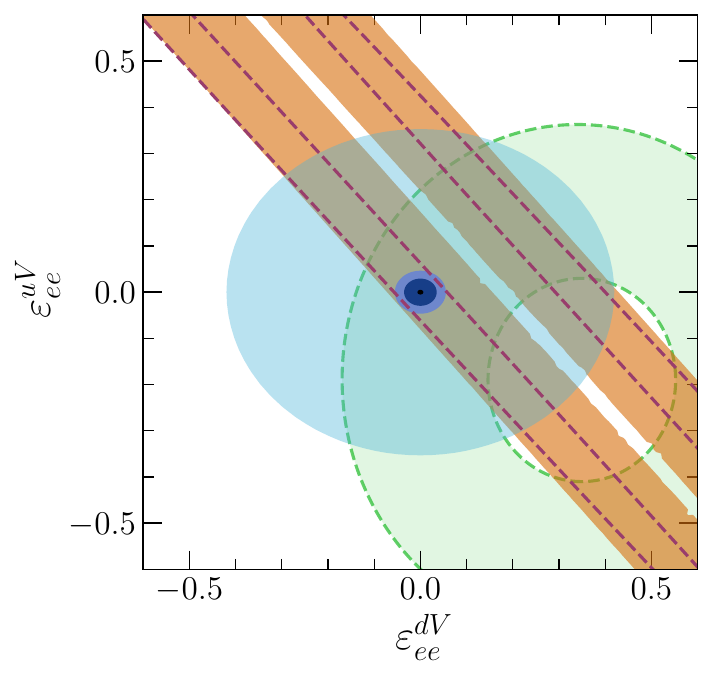}
\end{minipage}
\begin{minipage}[t]{0.48\textwidth}
  \includegraphics[width=\linewidth]{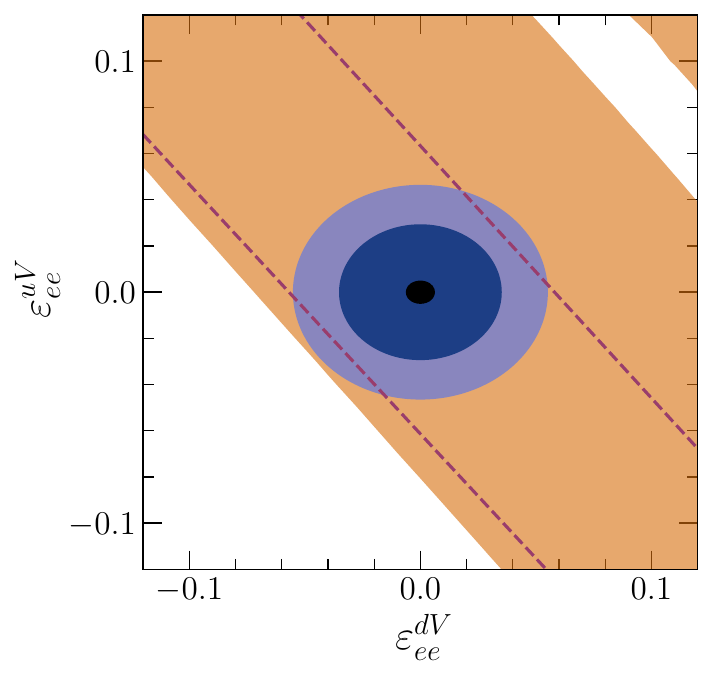}
\end{minipage}%

\begin{minipage}[t]{0.48\textwidth}
  \includegraphics[width=\linewidth]{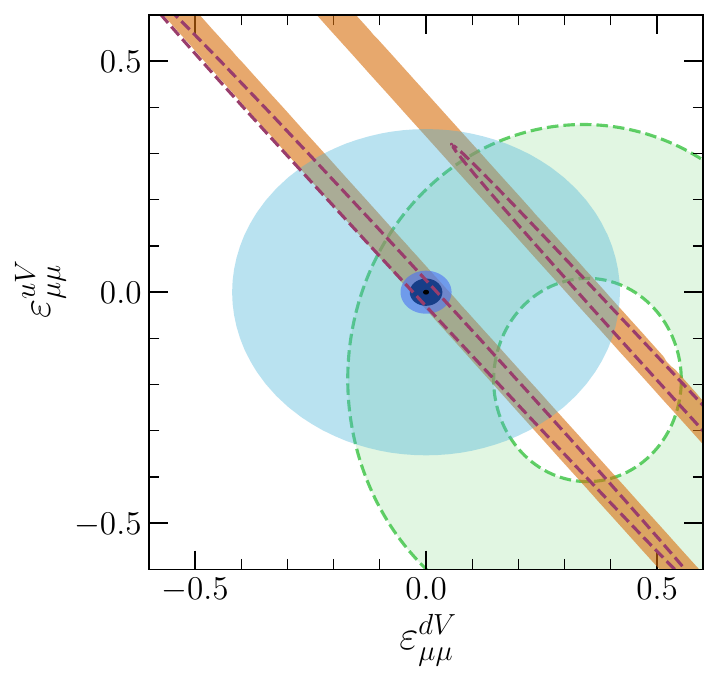}
\end{minipage}%
\begin{minipage}[t]{0.48\textwidth}
  \includegraphics[width=\linewidth]{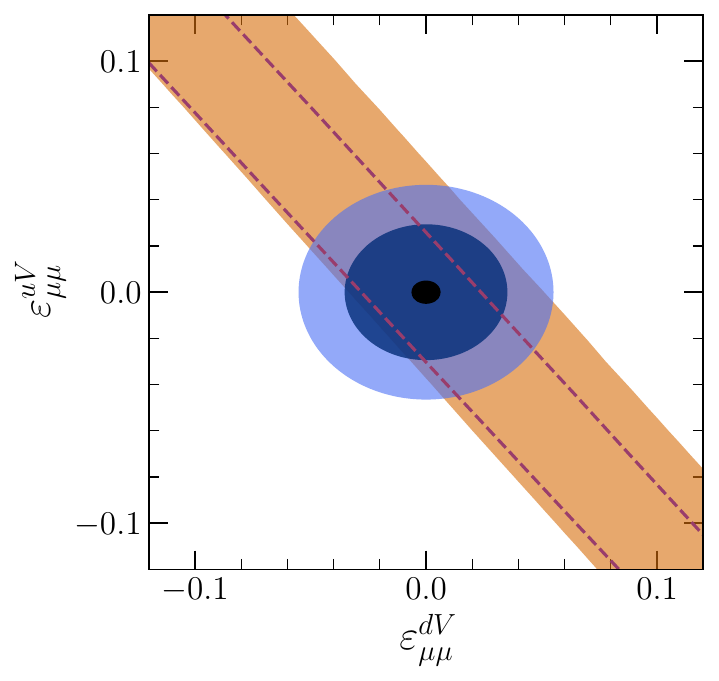}
\end{minipage}%

\begin{minipage}[t]{0.48\textwidth}
  \includegraphics[width=\linewidth]{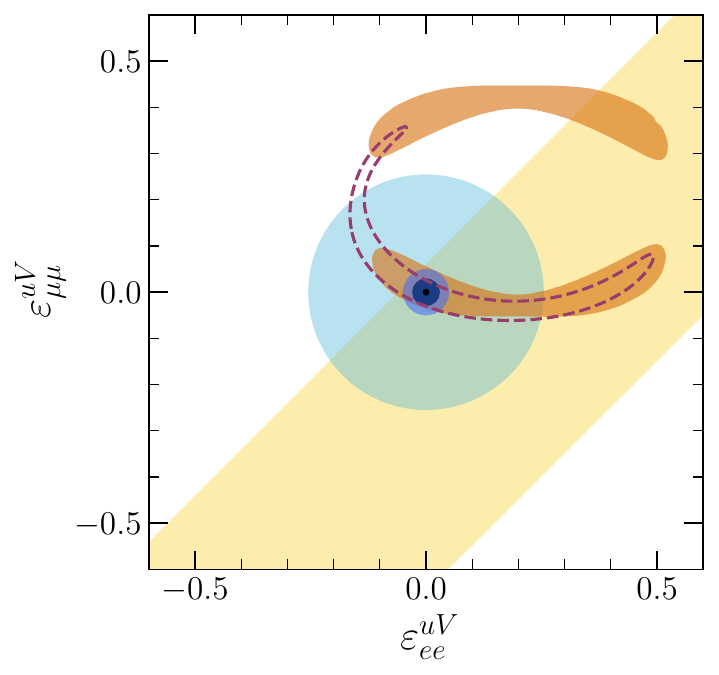}
\end{minipage}%
\begin{minipage}[t]{0.48\textwidth}
  \includegraphics[width=\linewidth]{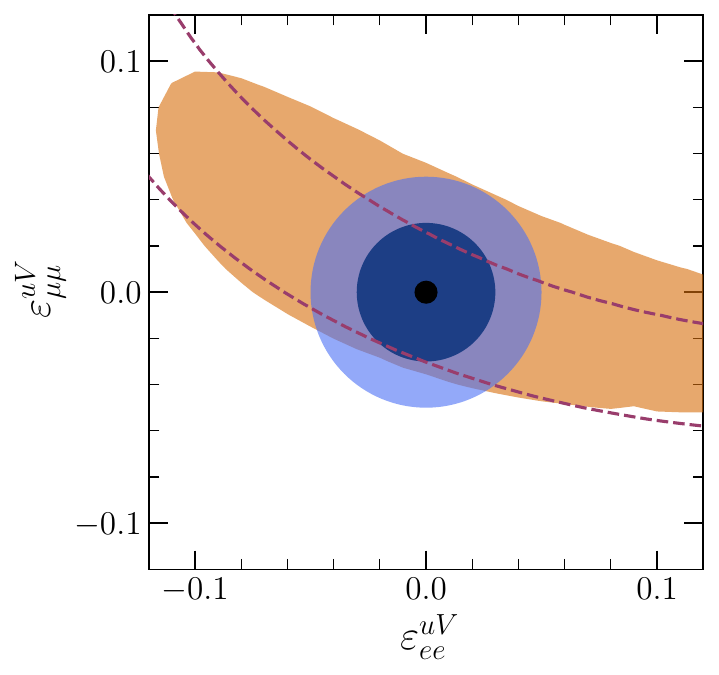}
\end{minipage}%
  \caption{Allowed parameter space at 95\% C.L. from COHERENT and LHC data in the $(\varepsilon^{dV}_{ee},\varepsilon_{ee}^{uV})$ (top row), $(\varepsilon^{dV}_{\mu\mu},\varepsilon_{\mu\mu}^{uV})$ (central row), and $(\varepsilon_{ee}^{uV},\varepsilon^{uV}_{\mu\mu})$ (bottom row) planes. The allowed parameter space from COHERENT CsI+LAr data is shown as a light orange region, while LHC results appear as blue regions, with shades from light to dark corresponding to $m_{Z'}$ values of 70 (light blue), 80 (blue), 100 (dark blue), and 300 (black) GeV. Future projections from COHERENT Cryo-CsI+NaI are represented by a dashed line boundary, while FASER ones are depicted as a light green region. The yellow band in the bottom left panel shows the oscillation data bounds taken from \cite{Coloma:2017ncl}. The right panel in each row shows an enlarged area of the corresponding left panel.}
  \label{fig:lhcoherent}
\end{figure*}

Until now, we have studied simplified cases where neutrinos either couple to up and down quarks with the same strength or they couple only one quark type. In the following, we will consider a more general scenario assuming these NSI couplings are different. In Fig.~\ref{fig:lhcoherent}, we show the 95\% C.L. allowed regions for different NSI pairs, allowing for a different coupling regardless of the quark type and neutrino flavor. We depict the results for LHC and COHERENT in the planes $(\varepsilon^{dV}_{ee},\varepsilon_{ee}^{uV})$, $(\varepsilon^{dV}_{\mu\mu},\varepsilon_{\mu\mu}^{uV})$, and $(\varepsilon^{uV}_{ee},\varepsilon_{\mu\mu}^{uV})$, with all the other NSIs assumed to be zero in each case. 
In the left panel of the first row, we show the results in the plane $(\varepsilon^{dV}_{ee},\varepsilon_{ee}^{uV})$, where allowed regions from the LHC are shown as blue areas for four different choices of $m_{Z'}$ in units of GeV: 70 (light blue), 80 (blue), 100 (dark blue), and 300 (black).\footnote{Notice that for these choices, only monojet searches are relevant to constrain NSIs.} For clarity, the corresponding right panel shows an enlarged area of the same figure. On the other hand, the constraints from the combined analysis from CsI and LAr are shown as orange regions. The two separate bands in this case correspond to regions around two minima in the statistical analysis, which is in agreement with the results presented in Ref.~\cite{DeRomeri:2022twg}. In general, \cevns\ analyses have shown that NSI degeneracies can be partially broken solely with this process by combining data from detector materials with different proportions of protons to neutrons (see, for instance, Refs. \cite{DeRomeri:2022twg, Chatterjee:2022mmu, Chatterjee:2024vkd}). However, we can see that this can also be achieved, particularly within the context of simplified models, by considering LHC bounds. 
To see this, notice from the figure that LHC bounds strongly depend on the mass of the mediator $m_{Z'}$, with more stringent limits as $m_{Z'}$ increases.
For $m_{Z'} = 70$ GeV, we see that the second band is not entirely excluded by LHC data, but it significantly reduces the allowed parameter space. In contrast, we see that for $m_{Z'}$ above 80 GeV, the second band of COHERENT is entirely excluded to be consistent with LHC data.

For completeness, we show as a dashed contour on the same figures the 95\% C.L. expected sensitivity from a combined analysis of two future COHERENT proposals: a cryogenic CsI (Cryo-CsI)~\cite{COHERENT:2023sol} and a sodium iodide (NaI) detector~\cite{Hedges:2021mra}. These regions were obtained by following the procedure in Ref.~\cite{Chatterjee:2024vkd}, where they were computed at a 90\% C.L. Regarding LHC searches, we also show as a green region the prospects from the forward detector, FASER$\nu$~\cite{Escrihuela:2023sfb}, which is now detecting neutrinos and could shed light on NSI couplings in the near future~\cite{FASER:2023zcr}. 
As we can see from the plots in the first row, the future prospects from Cryo-CsI+NaI can narrow the current bounds coming from CsI+LAr data. However, the degeneracy of the second band is still not broken by solely considering \cevns\ results, and it can only be excluded by LHC data. We notice that, for masses around $m_{Z'} = $ 80 GeV, LHC data and future \cevns\ experiments' sensitivity will be at the same level, still constraining different regions in the parameter space, while for $m_{Z'} \ge 100$ GeV, the limits are completely dominated by LHC data, since its corresponding regions are entirely contained within the COHERENT bands. Future prospects from FASER$\nu$ can also reduce the allowed parameter space in the two bands. However, we see that it is not expected to break the degeneracy between them at the same level as monojet searches. In general, we can see that the combination of LHC results with \cevns\, brings a complementarity that may allow one to break degeneracies in the parameter space when considering simplified models.

Moving to the other considered NSI pairs, in the central row of Fig.~\ref{fig:lhcoherent}, we show the results for the parameter space $(\varepsilon^{dV}_{\mu\mu},\varepsilon_{\mu\mu}^{uV})$, again with the right panel corresponding to an enlarged area of the left panel, under the same color code as in the previous case. We first notice that the constraining power from COHERENT is more robust than that for $(\varepsilon^{dV}_{ee},\varepsilon_{ee}^{uV})$, with the combined analysis of current CsI+LAr data resulting in two thinner separate bands (orange). We see that, again, for $m_{Z'}$ = 70 GeV (light blue) the second band is not excluded by LHC data. However, we notice a difference for $m_{Z'}$ = 80 GeV (blue) in a sense that, in contrast to the electron case, this region is not entirely contained in the allowed one by COHERENT. In this case, bounds from LHC start to be dominant from $m_{Z'} =$ 100 GeV (dark blue). Comparing now to the future COHERENT proposals of Cryo-CsI+NaI (dashed contour), it is expected that LHC bounds will be dominant only for mediator masses $m_{Z'}$ larger than 100 GeV. Finally, we present in the bottom row of Fig.~\ref{fig:lhcoherent} the results for two neutrino flavors coupling only to the up quark, that is, in the plane $(\varepsilon_{ee}^{uV},\varepsilon_{\mu\mu}^{uV})$, with the right panel corresponding to an enlarged region of the left panel. For reference, here we also show in yellow the bounds obtained from oscillation data~\cite{Coloma:2017ncl}. As in the previous cases, current COHERENT results (orange) exhibit a degeneracy that can be broken by considering LHC data. Moreover, it is interesting to note that, in this case, LHC can exclude the entire second region for mediator masses as low as $m_{Z'} = 70$ GeV (light blue), and LHC bounds are dominant for $m_{Z'} > 80$ GeV. Interestingly, we note that future prospects in \cevns\ (dashed) will also break this degeneracy, and LHC searches will only be dominant for $m_{Z'} > 70$ GeV (dark blue).

From our results, we conclude that, within our considered simplified models, the complementarity between LHC and \cevns\ data is manifest. As collider data are equally sensitive to all the neutrino NSI parameters and depend on the mass of the mediator, it allows one to break degeneracies that appear in the \cevns\ results. On the other hand, \cevns\ data are insensitive to mediator masses, but have a limited capability of breaking degeneracies by itself. It is worth mentioning that scattering experiments at high energies like NuTeV~\cite{NuTeV:2001whx} can also give strong  constraints on NSI competitive with the considered LHC bounds; see, for instance, Ref.~\cite{Coloma:2017egw}. However, these are highly dependent on theoretical uncertainties coming from nuclear effects. As a final remark, it is important to mention that although we only tested vector couplings in the NSI sector, it is possible that new physics also encodes axial couplings. In that case, the complementarity between experiments is of larger impact since the LHC is able to probe axial currents while COHERENT is not. These results show the complementarity between LHC and \cevns\ for disentangling new physics interactions in the neutrino sector.

\subsection{Charged leptons}

If NSI parameters come from higher-dimensional operators involving lepton doublets such as in Eq.~\eqref{eq:6dim}, then both neutral and charged states can have the same coupling to quarks, and, therefore, can be described by the Lagrangian of Eq.~\eqref{eq:simnsileptons}. If that is the case, the new physics could manifest also as new interactions, mediated by the $Z'$, between charged leptons and quarks. These new interactions can be tested at the LHC through final states involving a pair of charged leptons. In this section, we will briefly explore this scenario, where NSI parameters can modify the interactions of charged leptons. Again, we make use of the software tool $Z^\prime$-explorer~\cite{Alvarez:2020yim,Lozano:2021zbu} to test the expected signal when considering NSIs, this time including the dielectron and dimuon analysis~\cite{CMS:2021ctt} at $\sqrt{s}=13$ TeV and an integrated luminosity of $\mathcal{L}=139$ fb$^{-1}$ together with ditau~\cite{ATLAS:2017eiz} at $\mathcal{L}=36$ fb$^{-1}$, in addition to the mono- and dijet analysis we used before.

\begin{figure}[ht]
	\centering
        \includegraphics[width=.49\textwidth]{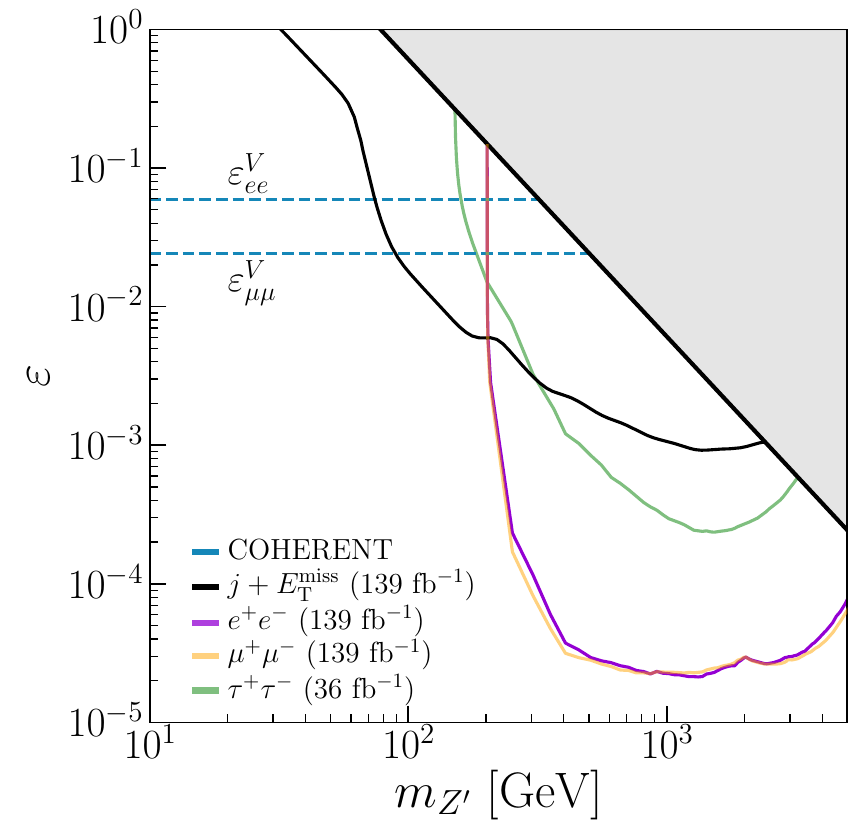}
        \caption{Constraints on the NSI parameter $\varepsilon$ as a function of $m_{Z'}$ from LHC and COHERENT data for the cases where the $Z'$ couples equally to up and down and couples equally to all lepton flavors. We display monojet (black), dielectron (violet), dimuon (yellow), and ditau (green) results and COHERENT bounds are depicted as horizontal lines. }
	\label{fig:cleptons}
\end{figure}

In Fig.~\ref{fig:cleptons} we show the results from the LHC analysis as a function of $m_{Z'}$, assuming that the $Z'$ couples equally to all lepton flavors. The different lines correspond to the bounds imposed by each of the LHC analysis; namely, monojet (black), dielectron (violet), dimuon (yellow), and ditau (green). We see that, for masses $m_{Z'}\lesssim 250$ GeV, the results are similar to those from Fig.~\ref{fig:epsilonmz}, with the slight difference that now the monojet search has a smaller constraining power. This effect is produced because allowing the $Z'$ to decay into charged leptons reduces the branching ratio of the $Z'$ decaying into neutrinos.
 However, the main difference with respect to the results in Fig.~\ref{fig:epsilonmz} is the presence of stringent bounds for masses $m_{Z'}\gtrsim 250$ GeV, which come from the charged lepton decays of the $Z'$. These searches are able to constrain values up to $\varepsilon\sim 10^{-5}$, with the most important searches coming from muons and electrons, while the tau search has less constraining power due to the low tau reconstruction and the low luminosity data ~\cite{ATLAS:2017eiz}. For reference, we show as horizontal lines in the figure the bounds obtained from COHERENT data, which are dominant for masses below 70 GeV.
\section{Conclusions}
\label{sec:conclusions}

We have explored the complementarity between COHERENT and LHC searches in testing neutrino NSIs. Given the different energy scales of these experiments, we have investigated the completion of the effective field theory that led to NSIs into a simplified model with vector-type currents. Under this formalism, COHERENT bounds are independent of the $Z'$ mass, while the LHC ones are largely dependent on it but blind to the neutrino flavor.

Among the considered scenarios, we first focused on the case where new physics only manifests in the neutrino sector, distinguishing two cases: one where neutrinos equally couple to up and down quarks, and another where they only couple to one quark type. For both cases, summarized in Fig. \ref{fig:epsilonmz}, we found that COHERENT bounds are more robust for $m_{Z'} < 70$ GeV, reaching a sensitivity of $\varepsilon\simeq 10^{-2}$. Above this range, LHC data are dominant and constrain up to $\varepsilon\simeq 10^{-3}$, assuming same coupling to up and down quarks and up to $\varepsilon\simeq 10^{-4}$ when only coupling to one quark type.

For a more general view, we have studied different cases allowing NSI couplings to be different. In particular, we performed this study for the $(\varepsilon^{dV}_{ee},\varepsilon_{ee}^{uV})$, $(\varepsilon^{dV}_{\mu\mu},\varepsilon_{\mu\mu}^{uV})$ and $(\varepsilon^{uV}_{ee},\varepsilon_{\mu\mu}^{uV})$ cases, summed up in Fig.~\ref{fig:lhcoherent}, finding that LHC data can break degeneracies present in COHERENT analysis for mediator masses above 80 GeV for the first two pairs and even from masses of 70 GeV for the parameter space $(\varepsilon^{uV}_{ee},\varepsilon_{\mu\mu}^{uV})$. For these scenarios, we have also compared our results with future prospects from COHERENT Cryo-CsI+NaI and FASER, finding that LHC will still be needed to break degeneracies within the context of simplified models.

Finally, we have studied the scenario allowing new physics for both the neutrino sector and charged leptons. In this case, we found that LHC searches have a strong constraining power for masses $m_{Z'}\geq 200$ GeV coming from the sensitivity of dilepton analysis at the LHC. However, for masses $m_{Z'}\leq 100$ GeV, the most robust constraints come from COHERENT.

Our results show that LHC and COHERENT results are complementary in order to unveil the nature of NSIs within a formalism of simplified models. On the one hand, LHC can determine the energy scale of new physics and may break degeneracies in COHERENT results, while COHERENT data would provide better information on the neutrino and quark flavor nature of new physics.

\section*{Acknowledgments}

We thank Jack D. Shergold for fruitful discussions regarding the EFT origin of the NSIs. This work has been supported by the Spanish Grants No. PID2023-147306NB-I00 and Severo Ochoa Excellence Grant No. CEX2023-001292-S (MCIU/AEI/10.13039/501100011033), both funded by Agencia Estatal de Investigación (AEI), and by Prometeo CIPROM/2021/054, funded by Generalitat Valenciana.  A. T. is supported by the Grant No. CIACIF/2022/159, funded by Generalitat Valenciana. G. S. G. acknowledges financial support by the Grant No. CIAPOS/2022/254, funded by Generalitat Valenciana.

\bibliography{bibliography} 

\end{document}